\documentclass[10pt,superscriptaddress,twocolumn,showkeys,aps,jcp]{revtex4-1}
\usepackage{epsfig,amsfonts}
\usepackage{mathrsfs}
\usepackage{bbold}
\usepackage{float}
\usepackage{graphicx}
\usepackage{bm}
\usepackage{amsmath} 
\usepackage{booktabs}

\usepackage{pstricks}
\usepackage{pst-solides3d}
\usepackage{pst-3d}
\usepackage{pst-3dplot}

\usepackage{pst-fr3d}
\usepackage{pst-gr3d}
\usepackage{pst-map3dII}
\usepackage{pst-ob3d}
\usepackage{pst-vue3d}

\usepackage{pstricks-add}

\def\be{\begin{equation}}
\def\ee{\end{equation}}
\def\bea{\begin{eqnarray}}
\def\eea{\end{eqnarray}}
\def\bpar{\left(\!\!\begin{array}}
\def\epar{\end{array}\!\!\right)}
\def\bdar{\left|\!\!\begin{array}}
\def\edar{\end{array}\!\!\right|}
\def\bar{\begin{array}}
\def\ear{\end{array}}

\def\nbOne{{\mathbb 1}}

\def\e{{\rm e}}

\def\bsigma{{\mbox{\boldmath $\sigma$}}}

\def\mat#1{{{{
	\bf#1}}}}
\def\H{{\mat H}}

\def\Un{{\textrm{\bf 1}}_{2\times 2}}

\def\<{\langle}
\def\>{\rangle}

\begin{document}

\title{	Continuum model for chiral induced spin selectivity in helical molecules}
\author{Ernesto Medina}
\affiliation{Centro de F\'isica, Instituto Venezolano de Investigaciones Cient\'ificas, 21827, Caracas, 1020 A, Venezuela.}
\affiliation{Groupe de Physique Statistique, Institut Jean Lamour, Universit\'e de Lorraine, 54506 Vandoeuvre-les-Nancy Cedex, France.}
\affiliation{Department of Chemistry and Biochemistry, Arizona State University, Tempe, AZ 85287, USA}
\author{Luis A. Gonz\'alez-Arraga}
\affiliation{Department of Chemistry and Biochemistry, Arizona State University, Tempe, AZ 85287, USA}
\author{Daniel Finkelstein-Shapiro}
\affiliation{Department of Chemistry and Biochemistry, Arizona State University, Tempe, AZ 85287, USA}
\author{Bertrand Berche}
\affiliation{Groupe de Physique Statistique, Institut Jean Lamour, Universit\'e de Lorraine, 54506 Vandoeuvre-les-Nancy Cedex, France.}
\affiliation{Centro de F\'isica, Instituto Venezolano de Investigaciones Cient\'ificas, 21827, Caracas, 1020 A, Venezuela.}
\author{Vladimiro Mujica}
\affiliation{Department of Chemistry and Biochemistry, Arizona State University, Tempe, AZ 85287, USA}

\begin{abstract}
A minimal model is exactly solved for electron spin transport on a helix. 
Electron transport is assumed to be supported by well oriented $p_z$ type orbitals 
on base molecules forming a staircase of definite chirality. In a tight binding interpretation, the SOC opens up an effective 
$\pi_z-\pi_z$ coupling via interbase  $p_{x,y}-p_z$ hopping, introducing spin coupled transport. The resulting continuum 
model spectrum shows two Kramers doublet transport channels with a gap proportional to the SOC. Each doubly degenerate channel
satisfies time reversal symmetry, nevertheless, a bias chooses a transport direction and thus selects for spin orientation.
The model predicts which spin orientation is selected depending on chirality and bias, changes in spin
preference as a function of input Fermi level and scattering suppression protected by the SO gap. We compute the spin current 
with a definite helicity and find it to be proportional to the torsion of the chiral structure
and the non-adiabatic Aharonov-Anandan phase. To describe room temperature transport
we assume that the total transmission is the result of a product of coherent steps limited by the coherence length. 

\end{abstract}
\pacs{\\ 73.23.-b  Electronic transport in 
	mesoscopic systems,\\
	85.75.-d Magnetoelectronics; spintronics: 
	devices exploiting spin polarized transport or integrated 
	magnetic fields,\\
	68.65.-k  Low-dimensional, mesoscopic, 
	nanoscale and other related systems: structure and nonelectronic 
	properties.
}

\maketitle

\section{Introduction}
Spin selectivity is an exciting prospect for many technological applications
that require spin imbalance as a resource for information storage and processing\cite{Zutic,ZuticNature}. Spin 
polarized currents, are traditionally derived from optical orientation\cite{Stevens} when
material-optic interphases are required or ferromagnets\cite{Silsbee} involving gross power sources to control.
Spin polarized currents derived by more subtle means are then desirable in the spin-torque driven writing 
on magnetic surfaces, valve devices and more generally in spin based information processing gates.

There is now well established experimental evidence that chiral molecules have the tantalizing
property of polarizing electrons scattered either through 2D films or driven through individual chiral molecules\cite{Naaman,Naaman2}.
These represent two limiting cases for transport: For 2D films, electrons are emitted from 
a metallic surface and are asymptotically free, having and energy above the film potential barrier. On the
other hand, for transport through a single molecule, electrons tunnel between metallic contacts driven
by a potential bias or a two terminal set up\cite{Naaman3,DominguezAdame}. 

The former, scattering experimental setup, is understood theoretically in terms of  multiple scattering through the chiral target in the presence spin-orbit coupling (SOC)\cite{Yeganeh,Medina1,Varela,DominguezAdame}. The theoretical model requires interference between scalar potential events and spin-coupled scattering.  Chiral structure and SOC define an energy window in which the electron spin is effectively polarized\cite{Medina1,Rosenberg} in the range of 1-10 eV. A higher energy limit is required for multiple coherent scattering and electron probing of the chiral structure, while a minimal energy is required for the SO coupling to be significant. The spin-orbit
coupling strengths required are compatible with those expected form the atomic carbon cores, in the
energy scale of the meV. Room temperature polarisation is proposed to be due to
incoherent superposition of short coherent length polarization\cite{Medina1}. 

On the other hand in the experiments performed by biasing a DNA molecule attached to
metallic contacts\cite{Naaman3} have been addressed theoretically by approaches based on 
quantum transport\cite{Gutierrez,DominguezAdame,Guo,Eremko}
in an chiral electric field generated by the charge distribution of the molecule, or helical potentials.
The electrons are considered to follow paths on the helix sampling the chiral electric field.
An important feature of the model is the relation between the field intensities
producing SOC and the electron mobility. The models predict a strong polarisation effect
in spite of the small SOC because the low mobility enhances the resident times of
electrons in the potential. Considerations of the limited coherence length due to room
temperature operations have been discussed in ref.\onlinecite{Guo}.

We aim here to theoretically describe the single molecule experiments in more detail:
For electrons with energies below the molecular barrier, a current is driven by a potential difference, through single DNA molecules
in an STM set up\cite{Naaman3}. The experimental results consist of a spin dependent barrier to tunnelling of electrons
through the chiral molecule. A preferred spin direction results in a larger current than the opposite spin
direction. There is a gap between barrier heights that is independent of the molecule
length and is thus proportional to the SO coupling. A striking characteristic
of the I-V curve is that it is anti-symmetric ($V\rightarrow  -V$), and the preferred spin filtering direction
in one bias direction is opposite to the preferred spin of the other\cite{Naaman3}. Finally, the
bias needed to produce the same spin polarised current increases with molecule length. 

Here we present a model for the tunnelling of electrons through a chiral molecule which can be solved exactly
explaining many experimental features. The
model includes the intrinsic SOC whose source are the atomic cores of the carbon atoms sitting on the bases
of the DNA structure (with a definite orientation in space). The strength of such a coupling for tunnelling electrons can be readily derived from a tight binding approach, and involves both the intrinsic atomic value and contributions from the coupling between nearest neighbour bases sites of the DNA molecule. We can anticipate the results for this model by performing a symmetry analysis: Both the SO coupling and the
helix break inversion symmetry, but time reversal symmetry is preserved. This implies that the spectrum
of the spin-orbit active helix should be composed of Kramers degenerate doublets, separated by
the effective spin-orbit coupling gap. The quantum numbers of the helix of definite chirality 
comprise the kinetic energy index, the rotation sense of the electron and its spin. Each Kramers doublet
preserves time reversal symmetry, so that they comprise both rotation and both spin quantum numbers. 
On choosing a bias direction, only the channel (one of the two states in the doublet) for that sense of
propagation is selected and has an associated spin. Thus, the bias breaks time reversal symmetry by
only populating one channel and {\it spin selectivity results}.  

This paper is organised as follows: We first derive the Hamiltonian for a continuous one dimensional helix of a
fixed number of turns and chirality, in the presence of spin-orbit coupling, whose source is a local atomic core electric field in the $z$ direction. Electrons are constrained to follow the helical path on the corresponding eigenchannels. Two energies are defined; that of the free electron problem and the SO energy whose ratio determines the adiabaticity of spin transport\cite{Frustaglia04}.
Following, we obtain the channel energies and the corresponding exact eigenfunctions
as a function of two quantum numbers (current direction and spin) and a chirality index.
We then show that the spectrum implies that there are always two doubly degenerate
levels and each degenerate pair combines time reversed states. Whenever one biased
direction is chosen, time reversal is broken and a preferred spin is filtered for that direction.
The opposite spin state in the same direction is at a higher energy so that it requires a
higher bias to be occupied, a feature that suppresses backscattering\cite{Naaman3}. The 
resulting spin selected transport is evidenced through the computation
of the spin current along the helix, leading to spin accumulation observed in the experiments.

The final section addresses the problem of the SOC strength, where we report a tight-binding result
which might address the strength of the spin-orbit energy gap observed. We also argue that the
tunnelling through the DNA molecule occurs through an incoherent sequence dividing the molecule
into coherence-length long segments to explain the scaling of the barrier potentials. We end
with the conclusions.

\label{sec:intro}

\section{Derivation of the intrinsic electric field Hamiltonian on a helix}
\begin{figure}
\vspace{0cm}
        \epsfxsize=8.0cm
        \begin{center}
        \mbox{\epsfbox{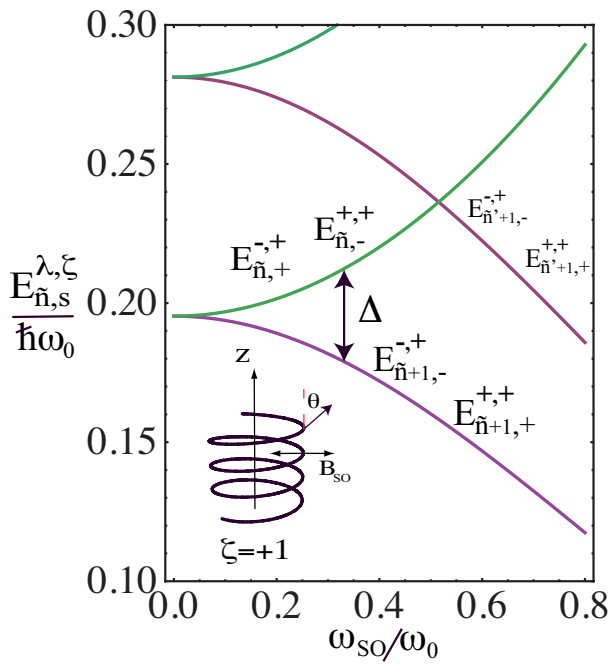}\qquad}
        \end{center}\vskip 0cm
        \caption{Energies of electrons on a counter-clockwise helix ($\zeta=+1$) as a function of the SO interaction strength $\omega_{\rm SO}$. The energy is in units  of $\hbar\omega_0$. Each level is doubly degenerate with
the index labels indicated. When $\omega_{\rm SO}=0$ each energy is fourfold degenerate as expected since the time inversion symmetry turns into independent space and spin inversion symmetries. $\Delta$ indicates the energy gap between Kramers degenerate states. The
inset shows the SO magnetic field and the orientation of the spin for an eigenstate of the spin-orbit coupled system.}
        \label{Fig1}  
\end{figure}

The model Hamiltonian we propose consists of an electron confined on a helix
of $N$ turns of radius $a$ and pitch $b$ as seen in Fig.\ref{Fig1}. An internal electric field, of atomic origin, is assumed to
exist as the source of the electron spin-orbit coupling. In an atomic model for the
helix, the SO interaction offers a $p_z$ to $p_{x,y}$ hopping route, first order in
the atomic intrinsic SO coupling and thus, at least in the meV energy range.
Assuming a motion strictly on the helix through a sequence of nearest neighbour states, 
we have the following Hamiltonian (see also ref.\cite{Tinoco})
\begin{eqnarray}
{\bf H}&=&{\bf H}_{\rm kinetic}+{\bf H}_{\rm SO}\notag \\
&=&\frac{1}{2m^*}(p_x^2+p_y^2+p_z^2)+(\alpha/\hbar)(p_x\bsigma_y-p_y\bsigma_x)
\end{eqnarray}
where the first term is pure kinetic energy, while the second term is the SO coupling
for an intrinsic electric field in the $z$ direction. The strength of the electric field is embedded in the
parameter $\alpha=e\hbar^2E/(4m^2c^2)$ which has dimensions of energy times a length. Nevertheless, 
this electric field is a nontrivial quantity to assess, since
it not only contains a measure of the field felt by the electrons in their excursion to their nuclei, but also a quasi-resonant
coupling contribution to the neighbouring states\cite{Guinea,Fabian}. We will discuss these contributions later.

Due to the symmetry of the problem it is preferable to work in cylindrical coordinates, and after
putting careful attention to hermiticity issues, well discussed in the literature\cite{Meijer,BercheHermitian}, we arrive at
\begin{equation}
{\bf H}=\frac{p_{\varphi}^2}{2m^*(a^2+b^2)}-\frac{\alpha a}{\hbar(a^2+b^2)}\bsigma_{\rho}p_{\varphi}+\frac{i \alpha a }{2(a^2+b^2)}\bsigma_{\varphi}
\label{Hamiltonian1}
\end{equation}
where $p_{\varphi}=m^*(a^2+b^2)\dot\varphi=-i\hbar \partial_{\varphi}$, $a$ is the helix radius, $b$ is the helix pitch, and 
$\bsigma_\rho=\bsigma_x\cos\varphi+\bsigma_y\sin\varphi$ and $\bsigma_\varphi=-\bsigma_x\sin\varphi+\bsigma_y\cos\varphi$.
The helix curvature is given by the ratio $\kappa=a/(a^2+b^2)$ and the torsion is $\tau=b/(a^2+b^2)$. The
momentum in the $z$ direction will then be $p_z=\tau p_{\varphi}$. $m^*$ is the electron effective mass, 
which assumes the carbon states form a narrow band of states. It is physically convenient\cite{Frustaglia04} to identify two distinct 
frequencies $\omega_0=\hbar/m^*(a^2+b^2)$ related to the free electron kinetic energy and $\omega_{\rm SO}=2\alpha a/\hbar(a^2+b^2)$ proportional to the helix curvature and the SO coupling. One can then simplify the
 the Hamiltonian in Eq.\ref{Hamiltonian1} into the simple quadratic form
\begin{equation}
H=\frac{\hbar \omega_0}{2}\left(i \partial_{\varphi}+\frac{\omega_{\rm SO}}{2\omega_0}\bsigma_{\rho}\right )^2.
\label{hamiltonianquadratic}
\end{equation}
The ansatz for the wave function for spinfull electrons is on a helix of $N$ turns with hard wall boundary conditions at the helix ends
\begin{equation}
\Psi_{n,s}^{\lambda,\zeta}=\e^{i \lambda (n/2N)\varphi}
\bpar{c}
A_{s}^{\lambda,\zeta}\e^{-i\varphi/2} \\ B_{s}^{\lambda,\zeta}\e^{i\varphi/2}
\epar,
\label{eigenfunction}
\end{equation}
where $\lambda=+1$($-1$) labels for the counter clockwise, (clockwise) electrons, $s=\pm 1$ labels the spin
and $\zeta=\pm 1$ labels the chirality of the helix.  Although the chiral index does not appear in the 
Hamiltonian, it chooses the $z$ direction of propagation for a particular $\lambda$ index.
Using the wave function ansatz one can derive the exact energy of the model as
\begin{equation}
E_{n,s}^{\lambda,\zeta}=\frac{\hbar \omega_0}{2}\left[\frac{n}{2N}-\frac{\zeta\lambda s}{2}\sqrt{1+\left(\frac{\omega_{\rm SO}}{\omega_0}\right)^2}~ \right]^2,
\label{energy}
\end{equation}
where $n$ is a positive integer valued index. The basis functions chosen in Eq.\ref{eigenfunction} are convenient when addressing biased conditions
for the system.

Note that left and right propagating electrons with the same $s$-index are not 
degenerate, but time reversal symmetry is satisfied i.e. 
$E_{n,s}^{\lambda,\zeta}=E_{n,-s}^{-\lambda,\zeta}$ (simultaneous change of $\lambda$ and $s$). This symmetry reflects the fact that the SO interaction is not symmetric under space inversion but preserves time reversal symmetry (simultaneous change of $\lambda$ and $s$) so 
we retain Kramers degeneracy. The chirality label $\zeta$ also reflects inversion asymmetry as changes in the chiral sign, at fixed $\lambda$ and $s$, change the energy. When the spin-orbit interaction is absent ($\alpha=0$), space and spin inversion symmetries are recovered (four fold degeneracy) once one combines $\tilde n=n/2N$ and $\tilde n+1$ labelled eigenvalues.

Figure \ref{Fig1} shows a sequence of levels starting from the ground state (Kramers doublets) at two successive values of $\tilde n$ 
along with their degenerate, at $\alpha=0$ partners with index $–+1$, as a function of the SO strength. The ordering of the 
levels are indicated according to  the spin orientation and sense of the current of the chirality $\zeta=+1$.

To obtain a physical intuition on the nature of the wave functions\cite{Frustaglia04,BercheHermitian} , we derive explicitly some of the coefficients in the 
ansatz put forward. We explicitly do the $\lambda=+,\zeta=+$ case.  Using equations (\ref{hamiltonianquadratic}) and 
(\ref{eigenfunction}) we find from the secular equation
\begin{equation}
B_{+,s}^{+}=\frac{\omega_0}{\omega_{\rm SO}}\left (\frac{s}{\cos{\theta}}-1\right )A_{+,s}^{+},
\end{equation}
where $\cos\theta=1/\sqrt{1+(\omega_{\rm SO}/\omega_0)^2}$. In order to conform 
to a normalised spinor we choose $A_{+,+}^+=\cos(\theta/2)$, and thus 
$B_{+,+}^+=\sin(\theta/2)$, so we have
\begin{equation}
\tan\theta=\frac{\omega_{\rm SO}}{\omega_0}.
\end{equation}
This angle results from the existence of a SO  ``magnetic field" \cite{Frustaglia04,Naaman4}~${\vec B}_{\rm SO}=-\alpha(\vec k\times \hat z)(2c/e)=(2 c \alpha/a\hbar e)\lambda|L|\hat\rho$, where $|L|$ is the angular momentum of the electron on the helix, and $\lambda$ gives the direction of the effective field depending on the rotation sense of the electron.

The choice for the second eigenfunction is $A_{+,-}^+=-\sin\theta/2$,
leading to the two eigenspinors
\begin{eqnarray}
\Psi_{n,+}^{+,+}&=&e^{i \tilde n\varphi}\bpar{c}
\cos\frac{\theta}{2}\e^{-i\varphi/2} \\ \sin\frac{\theta}{2}\e^{i\varphi/2},
\epar,\notag \\ 
\Psi_{n,-}^{+,+}&=&e^{i\tilde n\varphi}\bpar{c}
-\sin\frac{\theta}{2}\e^{-i\varphi/2} \\ \cos\frac{\theta}{2}\e^{i\varphi/2}
\epar,
\end{eqnarray}
The corresponding eigenfunctions for $\lambda=-$ are
\begin{eqnarray}
\Psi_{n,+}^{-,+}&=&e^{-i\tilde n\varphi}\bpar{c}
\cos\frac{\theta}{2}\e^{-i\varphi/2} \\ \sin\frac{\theta}{2}\e^{i\varphi/2}
\epar,\notag \\
\Psi_{n,-}^{-,+}&=&e^{-i\tilde n\varphi}\bpar{c}
-\sin\frac{\theta}{2}\e^{-i\varphi/2} \\ \cos\frac{\theta}{2}\e^{i\varphi/2}
\epar.
\end{eqnarray}
The angle $\theta$ here tells about the inclination of the spinor with respect to the vertical
$z$ axis\cite{Frustaglia04} (see Fig.\ref{Fig1}) and $\tilde n=n/2N$ with $n$ an integer. When the SO coupling is very strong, $\theta\rightarrow \pi/2$, and the spin is in the plane, while for very small SO coupling, the spin is aligned with the $z$ axis.
\begin{figure}[ht]
\begin{center}
\vspace{0cm}
\psscalebox{0.65} 
{
\begin{pspicture}(-1,-3)(13,7)

\uput[dr](1,5){\includegraphics[width=0.06\textwidth]{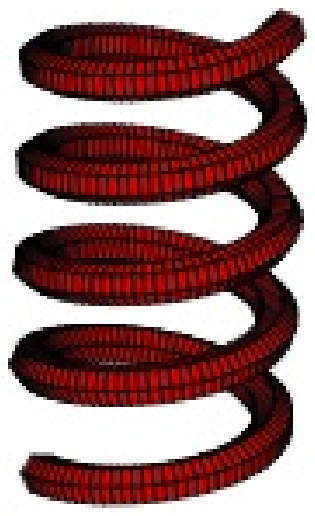}} 
\psline[linecolor=blue,linewidth=3pt,arrowscale=1]{->}(2.5 , 5)(2.5 , 3) 
\psline[linecolor=black,linewidth=3pt,arrowscale=1]{->}(2.8 , 3.5)(2.8 , 4.5)
\rput[bl](2.9,3.9){\Huge $s$} 
\rput[bl](1.2,5){\LARGE $\psi^{{\red -},+}_{\tilde n,{\red +}}$}

\uput[dr](3.5,5){\includegraphics[width=0.06\textwidth]{helix4.eps}}
\psline[linecolor=black,linewidth=3pt,arrowscale=1]{->}(5 , 3)(5 , 5)
\psline[linecolor=blue,linewidth=3pt,arrowscale=1]{->}(5.3 , 4.5)(5.3 , 3.5)
\rput[bl](3.8,5){\LARGE $\psi^{{\red +},+}_{\tilde n,{\red -}}$}
\rput[bl](2.8,2){\huge $E_>$}

\uput[dr](7,5){\includegraphics[width=0.06\textwidth]{helix4.eps}}
\psline[linecolor=black,linewidth=3pt,arrowscale=1]{->}(8.5 , 3)(8.5 , 5)
\psline[linecolor=black,linewidth=3pt,arrowscale=1]{->}(8.8 , 3.5)(8.8 , 4.5)
\rput[bl](6.9,5){\LARGE $\psi^{{\red +},+}_{\tilde n+1,{\red +}}$}

\uput[dr](9.5,5){\includegraphics[width=0.06\textwidth]{helix4.eps}}
\psline[linecolor=blue,linewidth=3pt,arrowscale=1]{->}(11 , 5)(11 , 3)
\psline[linecolor=blue,linewidth=3pt,arrowscale=1]{->}(11.3 , 4.5)(11.3 , 3.5)
\rput[bl](9.3,5){\LARGE $\psi^{{\red -},+}_{\tilde n+1,{\red -}}$}
\rput[bl](8.8,2){\huge $E_<$}
\uput[dr]{0}(1,0){\includegraphics[width=0.06\textwidth]{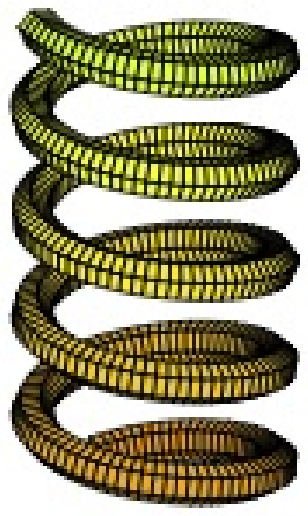}}
\psline[linecolor=blue,linewidth=3pt,arrowscale=1]{->}(2.5 , 0)(2.5 , -2)
\psline[linecolor=black,linewidth=3pt,arrowscale=1]{->}(2.8 , -1.5)(2.8 , -0.5)
\rput[bl](1.2,0){\LARGE $\psi^{{\red +},-}_{\tilde n,{\red +}}$}

\uput[dr]{0}(3.5,0){\includegraphics[width=0.06\textwidth]{helix3.eps}}
\psline[linecolor=black,linewidth=3pt,arrowscale=1]{->}(5 , -2)(5 , 0)
\psline[linecolor=blue,linewidth=3pt,arrowscale=1]{->}(5.3 , -0.5)(5.3 , -1.5)
\rput[bl](3.8,0){\LARGE $\psi^{{\red -},-}_{\tilde n,{\red -}}$}
\rput[bl](2.8,-3){\huge $E_>$}

\uput[dr]{0}(7,0){\includegraphics[width=0.06\textwidth]{helix3.eps}}
\psline[linecolor=black,linewidth=3pt,arrowscale=1]{->}(8.5 , -2)(8.5 , 0)
\psline[linecolor=black,linewidth=3pt,arrowscale=1]{->}(8.8 , -1.5)(8.8 , -0.5)
\rput[bl](6.9,0){\LARGE $\psi^{{\red -},-}_{\tilde n+1,{\red +}}$}

\uput[dr]{0}(9.5,0){\includegraphics[width=0.06\textwidth]{helix3.eps}}
\psline[linecolor=blue,linewidth=3pt,arrowscale=1]{->}(11 , 0)(11 , -2)
\psline[linecolor=blue,linewidth=3pt,arrowscale=1]{->}(11.3 , -0.5)(11.3 , -1.5)
\rput[bl](9.3,0){\LARGE $\psi^{{\red +},-}_{\tilde n+1,{\red -}}$}
\rput[bl](8.8,-3){\huge $E_<$}

\psline[linecolor=red,linewidth=2 pt,arrowscale=2]{->}(6,-2)(6,6)
\rput[bl]{90}(0.4,2.9){\LARGE $\zeta=+1$}
\rput[bl]{90}(0.4,-2){\LARGE $\zeta=-1$}
\rput[bl](6.3,1.5){\Huge $z$}

\end{pspicture}
}
\vspace{0cm}
\caption{Electron propagation direction coupled to spin orientation and the corresponding energies for the
        two possible chiralities of the helix. Each energy is doubly degenerate (Kramers doublet). If the electron propagates on a $\zeta=+1$ helix in the positive $z$ direction (as indicated), the lower energy state corresponds to the spin up state, while the spin down state has to pay the energy gap $\Delta$. } 
\label{Fig2}
\end{center}
\end{figure}
\section{Spin filtering and suppressed backscattering}

The rotation and spin eigenvalue corresponding to the wave functions is depicted in Fig.\ref{Fig2} for both chiral labels.
We have also evaluated the energies corresponding to the split doublets separated by the gap $\Delta$.
For fixed chirality, say $\zeta=+1$, there is a doubly degenerate (changing $s$ and $\lambda$ simultaneously) low 
energy configuration and a high energy configuration for each $\tilde n$.
\begin{eqnarray}
E_{>}&=&\frac{\hbar \omega_0}{2}\left [  \tilde n +\frac{1}{2}\sqrt{1+\left(\frac{\omega_{\rm SO}}{\omega_0}\right)^2}~ \right]^2\notag \\ 
E_{<}&=&\frac{\hbar \omega_0}{2}\left [  \tilde n+1 -\frac{1}{2}\sqrt{1+\left(\frac{\omega_{\rm SO}}{\omega_0}\right)^2}~ \right]^2 .
\end{eqnarray}
As shown in Fig.\ref{Fig2}, electrons propagating in the positive $z$ direction will have a lower energy if their
spin is $s=+1$. The same energy corresponds to their Kramers partner which propagates in the opposite direction.
In order to propagate a $s=-1$ state in the positive direction, we need to pay an energy price $\Delta$
 \begin{equation}
\Delta =\left (\tilde n+\frac{1}{2}\right )\frac{E_{\rm SO}^2}{2\hbar\omega_0},
\end{equation}
for $\omega_{\rm SO}/\omega_{\rm 0}<1$ and $E_{\rm SO}=\hbar\omega_{\rm SO}$. On the other hand, if the energy $\hbar\omega_0<\hbar\omega_{\rm SO}$ due to poor mobility of the electron on the helix (small $\omega_0$ due to large effective mass) then the energy gap is directly related to the
SO energy
\begin{equation}
\Delta = \left (\tilde n+\frac{1}{2}\right)E_{\rm SO},
\label{Gap}
\end{equation}
Thus, establishing a sense of electron propagation or bias will select a 
preferred spin direction (lowest energy). The selection of a propagation direction by applying a bias (as in $I-V$ experiments) or by local population imbalance, effectively breaks time reversal symmetry by selecting one of the time reversed partners and generating a net spin collected on the other end of the chiral molecule. Note also that scattering in the helix is suppressed by this gap if the spin is not concurrently flipped by the
scattering event\cite{Naaman4}. This is exactly the same mechanism for protected transport in chiral edge state in graphene\cite{KaneMele}.

To further connect with the experimental results, we note that if we change the bias direction for the same chirality label, the preferred 
spin is the opposite spin direction by choosing the other partner of the Kramers doublet of lowest energy. This change of selected spin is proven
in the experiments by the fact that spin injected back into the Ni magnet has to go where the DOS is higher, which is the
opposite spin state\cite{Naaman4}. 

The symmetry of the $I-V$ curve in experiments is directly related to the fact that the same degenerate energy
corresponds to the forward and backward bias, the change only being which of the two partners in the doublet is selected.
The barrier for electrons to be injected into the molecule depend on the workk function of the metal, and this 
is also very similar between Ni and Au ($\sim 5$ eV), so this is also a source of symmetry.


\section{Charge and spin currents}
Another way to pose spin selectivity is by evaluating the spin currents through the helix.
In a coherent regime one can assess spin transport in the helical molecule by computing the expectation value of the 
velocity operator $\vec J_{charge} =\Psi^{\dagger}e {\vec v}~\Psi$,
where $e$ is the electron charge and ${\vec {\bf v}}$ is the velocity operator. 
In the presence of the spin-orbit interaction, the velocity operator is not simply ${\vec p}~\nbOne/m$ (a diagonal matrix) as
there arises an additional anomalous velocity term. We start from the quantum mechanical definition of the 
velocity ${\vec  v}
= \frac{i}{\hbar}[\H,\vec r ]$. The azimuthal  velocity component $\dot\varphi$,
is then
\begin{equation}
{\vec{\mat v}}_{\varphi}=\frac{-i\hbar\partial_{\varphi}\Un-m a\alpha\bsigma_{\rho}}{m^*\sqrt{a^2+b^2}}.
\label{velocityoperator}
\end{equation}
The Hamiltonian takes a simple form when expressed in terms of the velocity: 
${\mat H}=\frac 12 m^*{\mat v}_{\varphi}^2$. This result is a manifestation of the possibility to write the SOC as a gauge
field\cite{EPL} and thus a gauge invariant velocity as defined in Eq.\ref{velocityoperator}.
Note that all the manifestations of spin filtering have been observed in biased setups. In the free electron case, 
photoelectrons are emitted from the metallic surface in contact with one end of the molecules, and travel in a preferred direction.
Also, in the localized regime, molecules are biased in a particular direction selecting the preferred sense of transport of the electrons.
This bias, as argued before, breaks time reversal symmetry and chooses between the states depicted in Fig.\ref{Fig2}.
We compute the charge currents for the biased system in the positive $z$ direction

\begin{equation}
J_{\tilde n+1,+}^{+,+}=-\frac{\hbar b\left [  \tilde n +1-\frac{1}{2}\sqrt{1+\left(\frac{\omega_{\rm SO}}{\omega_0}\right)^2}~ \right]}{m^*\sqrt{a^2+b^2}},
\end{equation}
versus the opposite spin state in the same propagation direction
\begin{equation}
J_{\tilde n,-}^{+,+}=-\frac{\hbar b\left [  \tilde n +\frac{1}{2}\sqrt{1+\left(\frac{\omega_{\rm SO}}{\omega_0}\right)^2}~ \right]}{m^*\sqrt{a^2+b^2}}.
\end{equation}
Thus, different spin states propagate at different velocities generating a net spin filtering effect. The net spin current 
vanishes for either SO zero or zero chirality ($b=0$). 

The difference is a net spin up current in the positive $z$ direction
\begin{eqnarray}
{\cal J}&=&J_{\tilde n+1,+}^{+,+}-J_{\tilde n,-}^{+,+}\notag \\
&=&\frac{\hbar b}{m^*\sqrt{a^2+b^2}}\left[ \sqrt{1+\left(\frac{\omega_{\rm SO}}{\omega_0}\right )^2}-1\right]
\end{eqnarray}

The longitudinal spin transport current is well defined and can be calculated as 
\begin{eqnarray}
{\cal J}_{\tilde n,z}^{+,+}&=&(\Psi_{\tilde n+1,+}^{+,+})^{\dagger}\frac{1}{2}\{{\mat v}_{\varphi},
{\mat s}^z\}\Psi_{\tilde n+1,+}^{+,+}\notag \\
&+&(\Psi_{\tilde n,-}^{+,+})^{\dagger}\frac{1}{2}\{{\mat v}_{\varphi},{\mat s}^z\}\Psi_{\tilde n,-}^{+,+}\notag\\
&=& \frac{\hbar^2 b}{2m^*\sqrt{a^2+b^2}}\left(\cos\theta -1\right).
\label{spincurrentzrashba}
\end{eqnarray}
The longitudinal spin current depends on the spin-orbit coupling through $\theta$. The spin filtered in the $z$
direction disappears when the SO coupling is zero ($\theta=0$). Again, the pitch dictates the strength of the
vertical spin current and both the pitch and the SOC must be present. Note that the factor $(1-\cos\theta)=\varphi_{\rm AA}/\pi$ is the 
non-adiabatic Aharonov-Anandan phase  found by Frustaglia and Richter in a detailed analyses of conductance through SO coupled rings\cite{Frustaglia04}. This phase offers a new insight into spin filtering of chiral molecules since in the strong SO limit it is related
to the Berry phase of the spin and for the general case it is controlled by non adiabatic spin precession. 

\section{Strength of the SOC and room temperature spin selectivity}
Recent experimental findings demonstrating spin selectivity have shown that the gap $\Delta$ in Eq.\ref{Gap}, is much larger
that expected from even the purely atomic contribution. We have attempted to explain this anomalous size of the SO gap
through a tight-binding approach\cite{Guinea,Fabian,Ando}. It is well known that the SOC can be enhanced three orders of magnitude from planar graphene to carbon nanotubes, just by bending the graphene sheets. 
In graphene, the intrinsic SOC is second order in
the atomic coupling ($\mu$eV second neighbour coupling) while it is linear in the atomic coupling for single walled nanotube
(meV from direct $p_x(i)-p_z(j)$ coupling\cite{Ando,Guinea} where $i,j$ are nearest neighbours). The latter
situation is also present in DNA if one regards the available orbitals to be on it bases, which are endowed with
available $p_z$ orbitals at such angle as to generate both $V_{pp\pi}$ and $V_{pp\sigma}$ couplings and linear
in atomic SOC. The derivation is beyond the scope of the present paper, but the results show nontrivial
features related to structural effects. 

We computed the effective coupling between DNA inter-base nearest neighbour $p_z(i)$ orbitals, assuming one
orbital per base, and a single helix. The same features should apply to the double helix. We found the
following form for the  SO coupling
\begin{equation}
V_{SO}\propto\frac{2\lambda(V_{pp}^{\pi}-V_{pp}^{\sigma})}{(\varepsilon_{2p}^{\pi}-\varepsilon_{2p}^{\sigma})\hbar},
\end{equation} 
The parameter $\lambda$ is the bare atomic SOC, 
$V_{pp}^{\pi}$ and $V_{pp}^{\sigma}$ are the Slater-Koster $p$ orbital couplings perpendicular and parallel
(respectively) to the line joining the carbon atoms of the model. $\varepsilon_{2p}^{\pi}-\varepsilon_{2p}^{\sigma}$ is the
difference in energy between the bonded $p_{x,y}$ in the plane of the base and the $p_z$ orbital perpendicular to
the plane. Note that the energy contrast in the denominator may be small, potentially increasing the SOC by a large amount. This is
especially true when interbase coupling is weak. Another way to express the denominator using lowest order perturbation theory is
by renormalizing the in plane coupling by the $s-p$ overlap; $\varepsilon_{2p}^{\pi}-\varepsilon_{2p}^{\sigma}\sim (V_{sp}^{\sigma})^2/(\varepsilon_{2p}-\varepsilon_{2s})$, where $\varepsilon_{2i}$ are the bare atomic values and $V_{sp}^{\sigma}$ is an interbase coupling that needs to be computed from a first principles study. 

The spin selectivity demonstrated by experiments is a room temperature phenomenon. Thus any
coherent mechanism must be limited by decoherence lengths\cite{Medina1,Guo}. The gap for degradation of spin selectivity found
in experiments  is very high, 0.5 eV (ref.\onlinecite{Naaman3}), compared to thermal effects at room temperature ~25 meV.
This gap prevents elastic backscattering and exponentially reduces inelastic scattering with the same
spin. A spin coupled scattering mechanism, nonetheless, could degrade the spin selectivity rapidly because
of the existence of the Kramers doublets as a backscattering channel (see Fig.\ref{Fig2}). 

If one proposes a decoherence length operating at room temperature, one can also suggest a mechanism which preserves spin selectivity analogous to that proposed in ref.\cite{Guo} and \cite{Medina1}, where relaxation of phase coherence
occurs within a few nanometers while spin selectivity is preserved. This mechanism entails
an exponential decay of the transmission with a decay rate\cite{Mujica} of $\beta=-(1/d_{\rm coh})\ln[t_{\rm eff}/(E_{\alpha}-(E_F-eV/2))]$
where $E_{\alpha}$ is the energy of the spin preffered channel, $t_{\rm eff}$ is the hopping integral between
sites separated by a coherence length $d_{oh}$, $E_F$ is the Fermi energy and $V$ the applied bias. 
Then increasing the number of $d_{\rm coh}$ by having longer molecules will make the input current
decrease exponentially, so and increased bias is needed to achieve the same output current. This
mechanism might be the source of the increased barrier for spin selectivity as the molecule is elongated.

 Again, there is no mechanism for the 
coupling to backscattering channels with the same spin (See Fig.\ref{Fig2}) so that there will
be degradation of spin current but not of spin polarization.

\section{Summary and Conclusions}

We have examined a model for chiral spin selectivity on a spin orbit active helix with $N$ turns. The
origin of the SO electric field comes from the atomic cores of the carbon atoms which provide,
through the $p_z$ orbitals, a narrow band for transport. The resulting channels of the
model helix are Kramers doublets involving opposite propagating and opposite spin
projections at the same energy. An applied bias or otherwise preferred transport direction will
then select a spin and effectively break time reversal symmetry (choosing one in a doublet pair), and transferring a particular
spin. The resulting spin current is the spin selectivity mechanism. The spectrum of the
model also insures suppressed backscattering by an energy gap controlled by the
SO energy and the inter-base coupling of the $p_z$ orbitals, as suggested by a tight
binding calculation. A form for the coupling is proposed, that could enhance the SOC in lowest order
perturbation theory.

There are a few predictions of the theory that could be tested experimentally:  1) The spin selected, after changing 
the chirality of the transport structure, seems to be the same although the state is different. 2) There should be changes in the 
strength of the SOC (as measured through the experimentally measured gap), by DNA deformations that
alter the inter-base hopping integrals. These could be induced by changing torsion (tugging on the molecule). The SOC, 
according to the tight-binding model, should increase as long as the molecule is not disrupted. 3) Current decay at
fixed bias when changing the molecule length. 4) Depending on the strength of the SOC and the length of the molecule, the
number of same-spin channels can increase making the selectivity more robust. 5) The suppressed backscattering
mechanism could be tested by changing DNA bases introducing random defects that do no couple to spin.
The spin selectivity should be robust against such inhomogeneities. 

The model here can be made more quantitative in many directions. The derivation of the Hamiltonian directly 
from the tight-binding descriptions is desirable so that one can properly account for the geometry
of the orbitals participating in transport. We have solved a variation of the problem more akin to the
geometry of the $p_z$ orbitals in carbon nanotubes, where they rotate on the outside of a cylinder.
Although the resulting Hamiltonian is different in detail the same physics described here follows. This is
expected from the symmetry arguments that result in Kramers doublets separated by a gap. 
Contemplating the double helix structure of DNA, could also bring about new interference effects
that might enhance or reduce spin selectivity\cite{Guo} and bring the model quantitatively closer to
the experiments.

Finally, incorporating 
the metallic contacts to the chiral molecule through the thiol groups, should introduce level broadening to the spectrum of the
molecule and determine the escape rate of electrons and possible charging phenomena.

\acknowledgments
The authors thank Ron Naaman, Mark Ratner and Abe Nitzan 
for enlightening discussions.  EM would like to gratefully acknowledge 
support from the Fulbright Foundation and IVIC Venezuela. 

\end{document}